\documentclass{svproc}
\usepackage{url}
\usepackage{subfig}
\usepackage{graphicx}
\usepackage{varioref}
\labelformat{equation}{\textup{(#1)}}
\usepackage[pageanchor=true,plainpages=false,pdfpagelabels,bookmarks,bookmarksnumbered]{hyperref}
\usepackage{amssymb}
\def\usenatbib{1}
\ifx\usenatbib\undefined
    \usepackage{cite}
\else
    \makeatletter
    \let\NAT@parse\undefined
    \makeatother
    \usepackage[numbers,sort]{natbib}
    \def\refname{References}

    \setcitestyle{citesep={], [}}
    \makeatletter
    \def\NAT@def@citea{\def\@citea{\NAT@separator}}%
    \makeatother
\fi

\graphicspath{ {./figures/} }

\let\orgautoref\autoref
\providecommand{\Autoref}
        {\def\equationautorefname{Equation}%
         \def\figureautorefname{Figure}%
         \def\subfigureautorefname{Figure}%
         \def\Itemautorefname{Item}%
         \def\tableautorefname{Table}%
         \def\exerciseautorefname{Exercise}%
         \def\starexerciseautorefname{Exercise}%
         \def\sectionautorefname{Section}%
         \def\subsectionautorefname{Section}%
         \def\subsubsectionautorefname{Section}%
         \def\chapterautorefname{Section}%
         \def\partautorefname{Part}%
         \orgautoref}

\providecommand{\Autorefs}
        {\def\equationautorefname{Equations}%
         \def\figureautorefname{Figures}%
         \def\subfigureautorefname{Figures}%
         \def\Itemautorefname{Items}%
         \def\tableautorefname{Tables}%
         \def\exerciseautorefname{Exercises}%
         \def\starexerciseautorefname{Exercises}%
         \def\sectionautorefname{Sections}%
         \def\subsectionautorefname{Sections}%
         \def\subsubsectionautorefname{Sections}%
         \def\chapterautorefname{Sections}%
         \def\partautorefname{Parts}%
         \orgautoref}

\renewcommand{\autoref}
        {\def\equationautorefname{Equation}%
         \def\figureautorefname{Fig.}%
         \def\subfigureautorefname{Fig.}%
         \def\Itemautorefname{item}%
         \def\tableautorefname{Table}%
         \def\exerciseautorefname{Exercise}%
         \def\starexerciseautorefname{Exercise}%
         \def\sectionautorefname{Section}%
         \def\subsectionautorefname{Section}%
         \def\subsubsectionautorefname{Section}%
         \def\chapterautorefname{Section}%
         \def\partautorefname{Part}%
         \orgautoref}

\providecommand{\autorefs}
        {\def\equationautorefname{Equations}%
         \def\figureautorefname{Fig.}%
         \def\subfigureautorefname{Fig.}%
         \def\Itemautorefname{items}%
         \def\tableautorefname{Tables}%
         \def\exerciseautorefname{Exercises}%
         \def\starexerciseautorefname{Exercises}%
         \def\sectionautorefname{Sections}%
         \def\subsectionautorefname{Sections}%
         \def\subsubsectionautorefname{Sections}%
         \def\chapterautorefname{Sections}%
         \def\partautorefname{Parts}%
         \orgautoref}

\begin{document}
\mainmatter              
%
\title{Beyond Tracking: Using Deep Learning to Discover Novel Interactions 
        in Biological Swarms}
\titlerunning{Beyond Tracking}  
%
\author{Taeyeong Choi\inst{1} \and Benjamin Pyenson\inst{2} \and
Juergen Liebig\inst{2} \and Theodore~P.~Pavlic\inst{2, 3, 4}}
\authorrunning{Taeyeong Choi et al.} 
%
\tocauthor{Taeyeong Choi, Benjamin Pyenson, Juergen Liebig,
Theodore P. Pavlic}
\institute{Lincoln Institute for Agri-food Technology, 
University of Lincoln, Riseholme Park, Lincoln, UK \\
\and
School of Life Sciences,  Social Insect Research Group, \\
Arizona State University, Tempe, AZ 85281, USA \\ 
\and 
School of Computing, Informatics, and Decision Systems Engineering, \\
Arizona State University, Tempe, AZ 85281, USA \\
\and
School of Sustainability, Arizona State University, Tempe, AZ 85281, USA \\
\email{tchoi@lincoln.ac.uk, \{bpyenson, jliebig, tpavlic\}@asu.edu}}

\maketitle              

\begin{abstract} 
Most deep-learning frameworks for understanding biological swarms
are designed to fit perceptive models of group behavior to
individual-level data (e.g., spatial coordinates of identified features
of individuals) that have been separately gathered from video
observations. Despite considerable advances in
automated tracking, these methods are still very
expensive or unreliable when tracking large numbers of animals
simultaneously. Moreover, this approach assumes that the human-chosen
features include sufficient features to explain important patterns in
collective behavior. To address these issues, we propose training deep
network models to predict system-level states directly from generic graphical 
features from the entire view, which can be relatively inexpensive to gather in a 
completely automated fashion. Because the resulting predictive models are not based
on human-understood predictors, we use explanatory modules (e.g.,
Grad-CAM) that combine information hidden in the latent variables of the
deep-network model with the video data itself to communicate to a human
observer which aspects of observed individual behaviors are most
informative in predicting group behavior. This represents an example of
augmented intelligence in behavioral ecology~-- knowledge co-creation in
a human--AI team. As proof of concept, we utilize a $20$-day video
recording of a colony of over $50$~\emph{Harpegnathos saltator} ants to
showcase that, without any individual annotations provided, a trained
model can generate an ``importance map'' across the video frames to
highlight regions of important behaviors, such as \emph{dueling} (which the AI
has no \emph{a~priori} knowledge of), that play a role in the resolution of
reproductive-hierarchy re-formation. Based on the empirical results, we
also discuss the potential use and current challenges to further develop
the proposed framework as a tool to discover behaviors that have not yet
been considered crucial to understand complex social dynamics within
biological collectives.

\end{abstract}
\section{Introduction}
\label{sec:intro}
%

Deep Convolutional Neural Networks~(DCNNs) have been widely adopted as 
the primary backbone of data-driven frameworks to solve complex problems in 
computer vision including object classification or
detection and recognition of human actions~\citep{SZ14, RF18, WXWQLTV16}. 
The nature of their
multi-layer structure has a powerful ability to automatically 
learn to identify key local features (e.g.,~edges) from raw pixels of
images 
and combine into more meaningful concepts (e.g.,~pointy ears) to produce a final 
prediction output (e.g.,~dog), as the data is processed from 
the lowest layer through the higher ones~\citep{GBCB16}. In fact, this may imply that 
if the target data contains global information of biological swarms, 
lower-level visual properties such as locations, motions, and interactions of the 
entities could automatically be identified throughout the hierarchical layers during 
the training process. However, deep learning in behavioral biology has
mostly been
limited to building perceptive models to localize particular body 
parts of each entity to generate another input 
to a subsequent analysis model to capture motional concepts of individuals and perform 
a prediction for the entire swarm based on 
them~\citep{BHMS18, RBHHD19, GCNLKCC19, NMCPBM19}.
\begin{figure}[t]
        \centering
        \includegraphics[width=.75\linewidth]{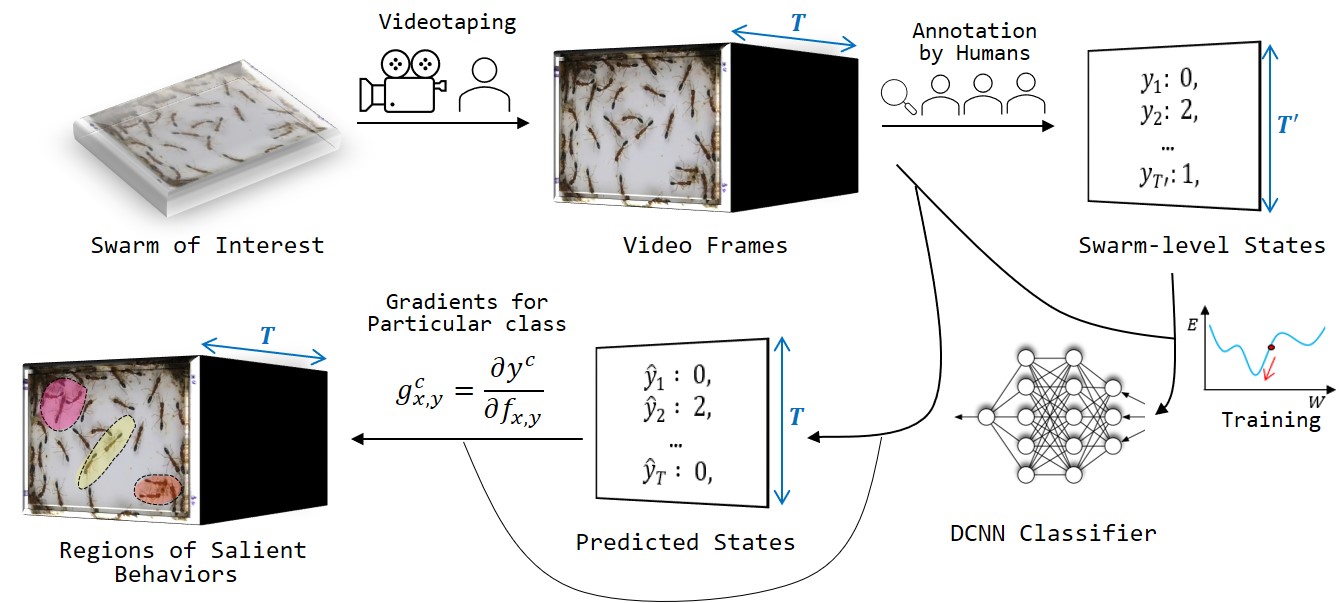}
        \caption
        {
        Proposed usage of DCNNs, trained to predict global state 
        of the swarm system from the entire view to later reveal key  
        local observations by using the gradient between the learned local feature 
        and the prediction output in the model.
        }
        \label{fig:Proposed}%
\end{figure}

There can be two main challenges in this approach: 1)~obtaining 
the individual feature labels can require a significant amount of human 
effort especially when a large group of system is examined, 
and 2)~the choice of features relies heavily on 
prior knowledge of human experts in the biological system. To address these 
issues, as visualized in~\autoref{fig:Proposed}, we here suggest training the 
deep-network models to predict system-level states directly from generic graphical 
features from the entire view, which can be relatively inexpensive to gather, and 
examine the salient behavioral regularities discovered in the trained
intermediate layers by using \emph{gradient}-based explanation modules
(e.g., Grad-CAM~\citep{SCDVPB17}). In other words, our proposal is to
make more use of the aforementioned potential of DCNN to automatically
discover fine-grained, individual-level motional patterns highly
associated with macroscopic swarm properties so that the predictive
model can later be queried about what these patterns are without being
constrained by prior knowledge from human experts.

\begin{figure}[t]
        \centering
        \begin{minipage}{.5\textwidth}
          \centering
          \includegraphics[width=.70\linewidth]{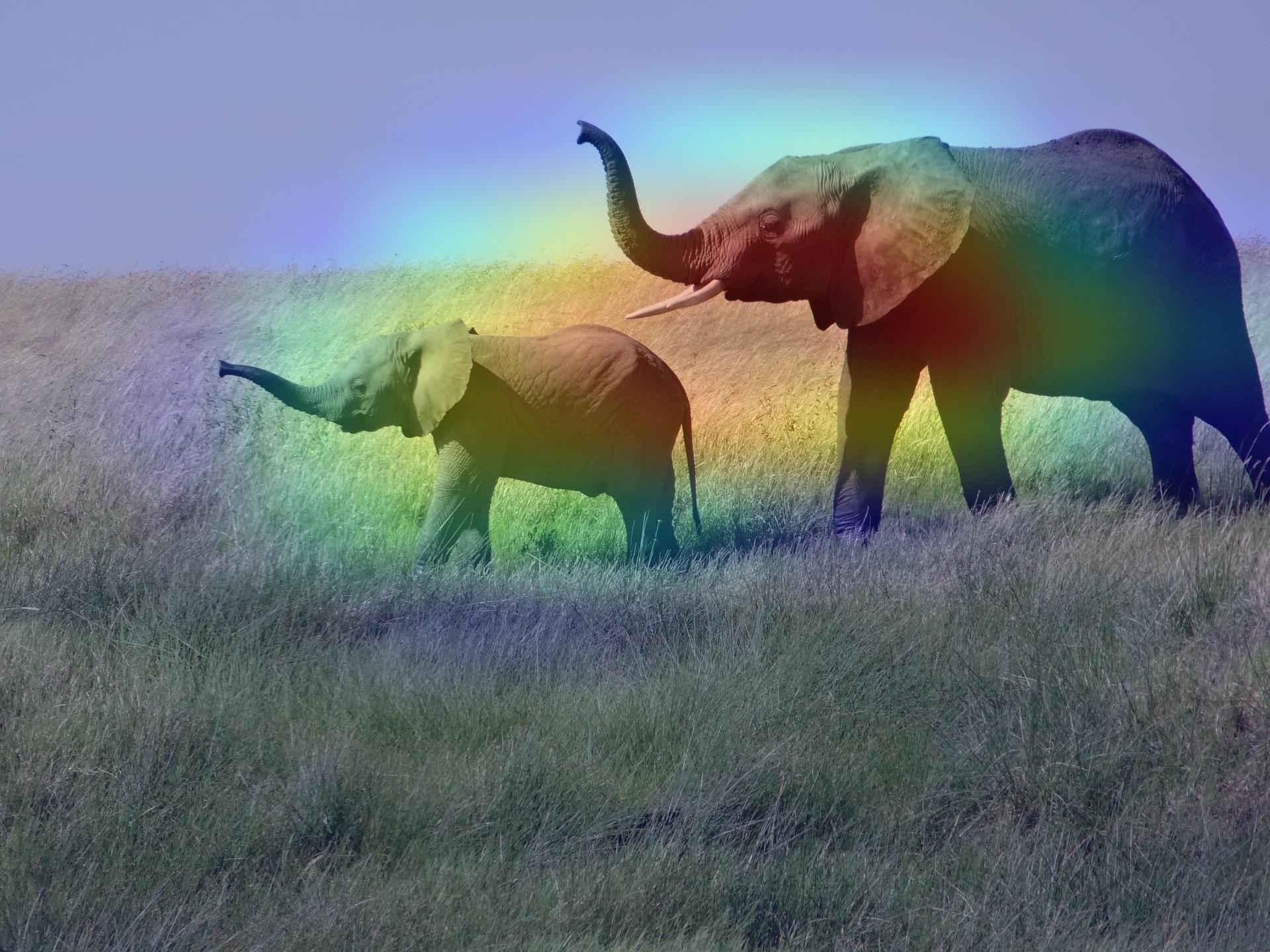}
          \captionsetup{width=.85\linewidth}
          \caption{Example of Grad-CAM in which the key regions 
                   are highlighted for class ``Elephant''\citep{KerasGradCam}. }
          \label{fig:grad_cam_example}
        \end{minipage}%
        \begin{minipage}{.5\textwidth}
          \centering
          \includegraphics[width=.70\linewidth]{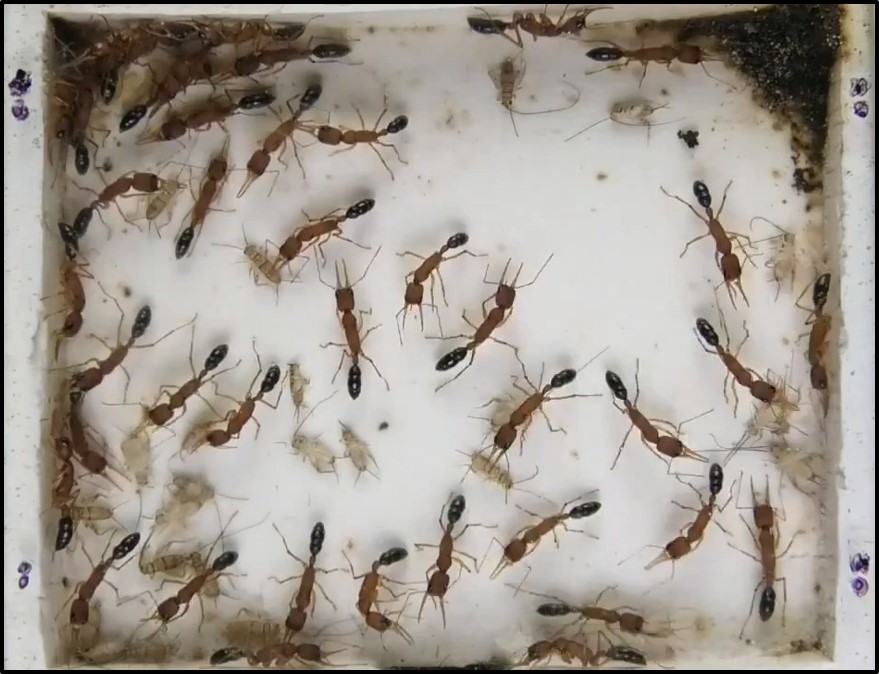}
          \captionsetup{width=.85\linewidth}
          \caption{Colony of $59$~\emph{H. saltator} as a testbed, with
          a foraging chamber accessed by the south tunnel.}
          \label{fig:colony_setting}
        \end{minipage}
\end{figure}

Specifically, in this paper, we propose the use of the explainable
module Grad-CAM~(\autoref{fig:grad_cam_example}) for biological
research. Extending our previous work~\citep{CPLP21}, we utilize a
$20$-day video recording of a colony of over $50$~\emph{Harpegnathos
saltator} ants to demonstrate that without any individual annotations
provided as input, the trained model can classify social stability of
colonies while also generating an ``importance map'' across video frames
to selectively highlight regions of interactions (e.g., \emph{dueling})
as potentially important drivers of colony state.
%

\section{Proposed Framework}
\label{sec:framework}
%


Rather than training on small-scale features of individuals in videos,
our approach trains a DCNN to predict coarse-grained, large-scale
labels~($y$) from representations of generic features from video data.
Any discrete, large-scale property can be used, such as whether a
crowd~\citep{MOS09} is about to riot. We use hierarchy state $y \in
\{Stable, Unstable\}$ for a \emph{H.~saltator} colony~\citep{CPLP21}.
Our $n$-layer classifier consists of $m$~two-dimensional convolutional
layers~$\phi_{1\leq \ell \leq m}$ followed by other types~$\psi_{m+1
\leq \ell' \leq n}$, such as recurrent or fully-connected layers.
Convolutional layers are used as feature extractors in this architecture
since each output~$f_{ij}$ at~$\phi_{\ell}$ can compactly encode 
the local observation in a larger region (``receptive field'') at previous
layers~$\phi_{\ell''<\ell}$; i.e., a change in $f_{ij}$ can imply 
the amplification or decrease of the motion pattern observed 
in the corresponding region.



For explanation of what visual regions are most important to the
predictive model,
Grad-CAM~\citep{SCDVPB17} is employed on $K$ two-dimensional output feature maps, 
each denoted as~$f^k \in \mathbb{R}^{h \times w}$, at a convolutional 
layer~$\phi_\ell$ to finally calculate the ``importance map''~$M^c$ over 
the original input for a particular class~$c$. In the technical aspect, 
$\phi_\ell$ can be an arbitrary layer satisfying $\ell \in \{1, 2, ..., m\}$, 
but the layer~$\phi_\ell$ close to~$\phi_m$ is typically chosen to access more 
abstract features with wider receptive fields than the ones available at lower 
layers~$\phi_{\ell'<\ell}$. For brevity, we denote $\phi$ to be the chosen 
convolutional layer in the following descriptions.

To generate the importance map $M^{c}$, we first obtain the 
gradient~$g^c$ of the output~$y^c$ with respect to each 
feature map~$f^k$ from $\phi$, i.e.,~$g_{ij}^{c}={\partial
y^{c}}/{\partial f^{k}_{ij}}$. Therefore, $g_{ij}^{c}>0$ implies that
enhancing the observational pattern encoded by $f_{ij}^{k}$ increases
the predicted likelihood of class~$c$~-- the discovered pattern is
``salient'' for class~$c$~-- and $g_{ij}^{c}\leq0$ implies that the
observation is considered irrelevant to the prediction of class~$c$.
Then, for each feature map~$f^k$, Grad-CAM then uses this quantity to gain the averaged
importance~$a^{c}_{k} = (1/Z) \sum_{i}\sum_{j} g_{ij}^c$ (where $Z$ is a
normalization constant).
%
Finally, the importance 
map~$M^{c}$ is computed by the weighted summation of feature
maps:
\begin{equation}
        M^{c} = \Gamma \bigg( \sum_{k} a_{k}^{c} \odot f^{k} \bigg)
\end{equation}
where $\odot$ is the element-wise multiplication, and $\Gamma(a)=a$ for
$a>0$ and $\Gamma(a)=0$ otherwise.
In~\autoref{sec:results}, 
we also introduce a more restrictive $\Gamma'$ that gates only
the top $5\%$ values so as to strictly verify whether key behaviors are 
effectively highlighted with the highest level of confidence. Also, 
$M^{c}$ can be spatially upsampled to fit the original image of a desired 
size for visualization purpose.

\section{Testbed Design with \emph{H. saltator}}
\label{sec:h_saltator}

As in~\citep{CPLP21}, a colony of \emph{H.~saltator} is utilized as a
testbed to validate whether our proposed framework can reveal salient
behavioral patterns. A conspicuous ``unstable'' state can be induced in
this system through the removal of identified egg layers
(``gamergates'')~\citep{PC85} that triggers a hierarchy reformation
process. During this process, aggressive interactions such as
\emph{dueling}~\citep{SPSHPL16} 
can be readily observed for several weeks until several mated workers activate
their ovaries and start to lay eggs as the new gamergates, causing the
colony to return to its nominal stable state~\citep{LPH99}. We apply
our framework to this system by building a binary-state classifier on
the stability of the colony. We use the resulting deep-network model to
identify important behaviors of interest and validate whether
\emph{dueling}~(\autoref{fig:rgbs_with_duel}) is discovered without
\emph{a~priori} knowledge of it. Other behaviors identified
by the system may then warrant further investigation by human
researchers.



\subsection{Video Data from Colonies Undergoing Stabilization}
\label{sec:video_recording}
\begin{figure}[t]
        \centering
        \subfloat[]{\label{fig:rgbs_with_duel}%
        \includegraphics[width=.45\linewidth]{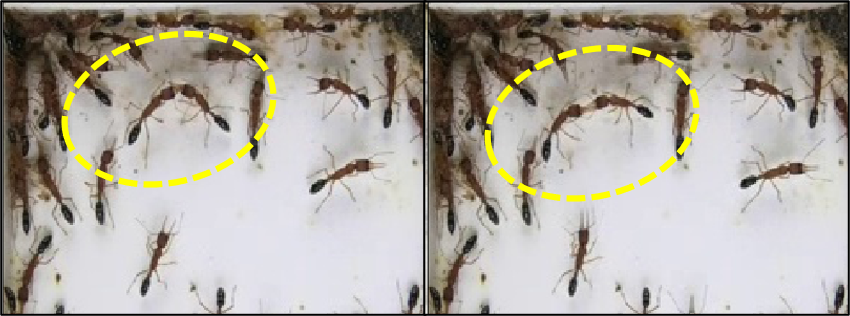}}
        \qquad
        \subfloat[]{\label{fig:optflow}%
        \includegraphics[width=.45\linewidth]{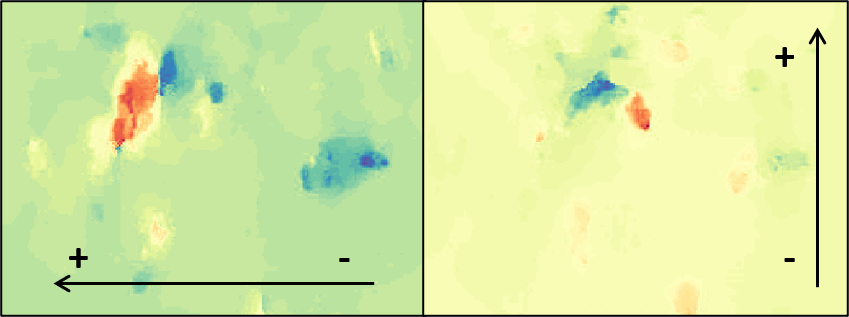}}
        \caption
        {
        \protect\subref{fig:rgbs_with_duel} Example of two consecutive RGB frames
        cropped around a \emph{dueling} interaction in yellow circle for 
        visibility; 
        \protect\subref{fig:optflow} Horizontal and vertical optical 
        flow vectors generated from \protect\subref{fig:rgbs_with_duel}, in each 
        of which red (blue) are the regions of movement in the positive 
        (negative) direction along the corresponding axis. 
        }
        \label{fig:optflow_example}
\end{figure}

As shown in~\autoref{fig:colony_setting}, each 20-day video was taken
with an overhead camera to observe $59$~\emph{H.~saltator} ants in
plaster nests covered with glass. Due to a foraging chamber outside the
view of the camera, not all ants are necessarily visible at all times,
and some paralyzed crickets can be carried into the view. We
artificially disturbed the reproductive hierarchy by removing all four
preidentified gamergates after the second day of recording and further observed 
the process of hierarchy reformation until aggressive interactions almost
disappeared in the last several days. Therefore, the video frames of the 
first $2$~days are annotated with $y=Stable$, while the later ones of $18$~days 
are all with $y=Unstable$.

We follow the preprocessing
method in~\citep{CPLP21} to extract from consecutive frames their optical flow, 
for which a pair of vectors encodes the horizontal and vertical transient 
movements from the input sequence 
(e.g., \autoref{fig:optflow_example})~\citep{MOS09}. 
Two optical flows in spatial resolution of $64 \times 64$
were computed every two minutes to use as an input~$x$ to the model, 
as each was obtained from two consecutive RGB frames $0.5$~seconds apart in times. 
More details of the dataset are available 
online\footnote{\url{https://github.com/ctyeong/OpticalFlows_HsAnts}}.

\subsection{Deployed DCNNs with Grad-CAM}
\label{sec:dcnns}

We use a classifier from our previous work~\citep{CPLP21} for the
\emph{one-class classification} task of predicting colony state. That
colony-state classifier has an overall performance of $0.786$ in the
Area Under the Curve~(AUC) of the Receiver Operating
Characteristic~(ROC) while only taking two consecutive optical flows as
input. Moreover, colony-state predictions during the early period of 
first $6$~days after the reproductive hierarchy is disturbed have 
higher AUC scores than $0.900$ in average~\citep{CPLP21}, 
indicating that the micro-scale graphical features identified by the 
deep network may be strong predictors of macro-scale state dynamics.

More specifically, the classifier we use has four $2$D convolutional
layers~$\phi_{1:4}$ with $2$D max pooling between consecutive layers,
and 
six other types of layers~$\psi_{5:10}$ follow to produce the estimated 
likelihood of unstable colony state. As described in~\autoref{sec:framework}, we
then employ Grad-CAM on the feature maps from $\phi_4$.
For each generated importance map~$M^c$, \emph{bicubic} interpolation is
applied to match the size of the frame image to overlay.

\section{Results and Model Validation}
\label{sec:results}
\begin{figure}[t]
  \centering
        \subfloat[]{\label{fig:centers}
        \includegraphics[width=.4\linewidth]{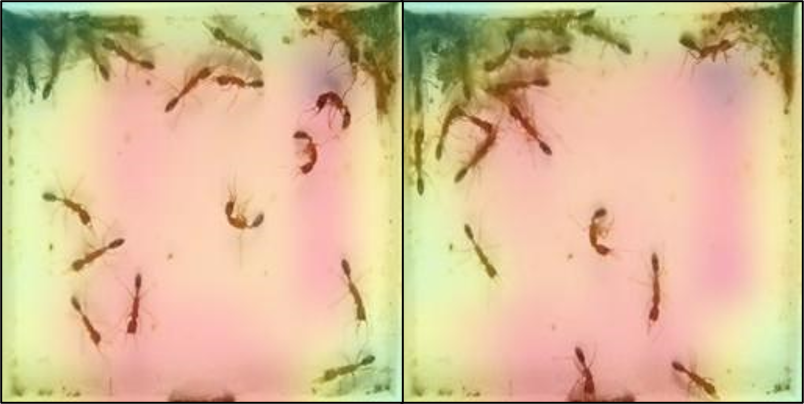}}
        \qquad
        \subfloat[]{\label{fig:grad_cam_duel_small1}
        \includegraphics[width=.4\linewidth]{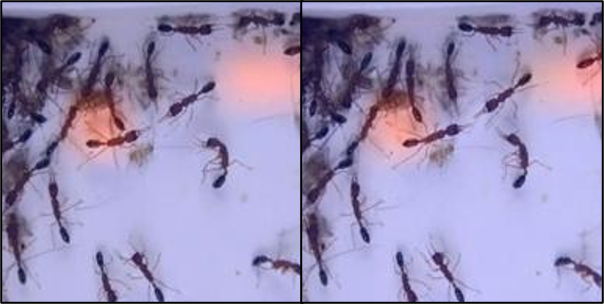}}
        \\
	\subfloat[]{\label{fig:grad_cam_duel_small3}
        \includegraphics[width=.4\linewidth]{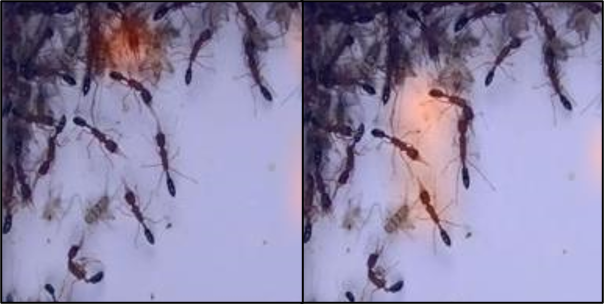}}
        \qquad
        \subfloat[]{\label{fig:grad_cam_duel_small4}
        \includegraphics[width=.4\linewidth]{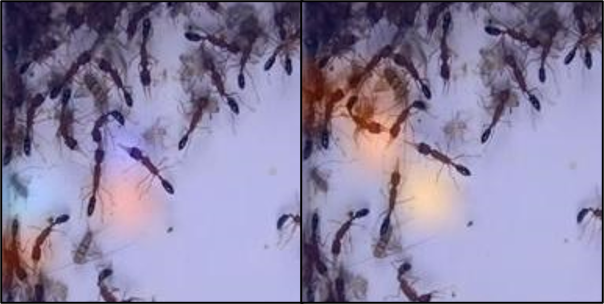}}
	\caption
	{
		\protect\subref{fig:centers}: Heatmaps from 
                Grad-CAM at two arbitrary times; 
		\protect\subref{fig:grad_cam_duel_small1},
                \protect\subref{fig:grad_cam_duel_small3},
                \protect\subref{fig:grad_cam_duel_small4}: 
                Three \emph{dueling} examples captured by the top 
                $5\%$~impactful regions of red. Each pair shows two consecutive 
                frames cropped around the interaction for clarity.
    }
    \label{fig:grad_cam_duel_smalls}
\end{figure}

As discussed in~\autoref{sec:h_saltator}, we validate our approach by
confirming that \emph{dueling} behavior between ants is identified by
the AI as strongly related to the unstable colony state.
A model that can detect \emph{dueling} with no prior knowledge of the behavior
may identify other behavioral patterns that warrant further
investigation.
%
%
\begin{figure}[t]
        \centering
        \subfloat[]{\label{fig:duel_non_duel1}
        \includegraphics[width=.4\linewidth]{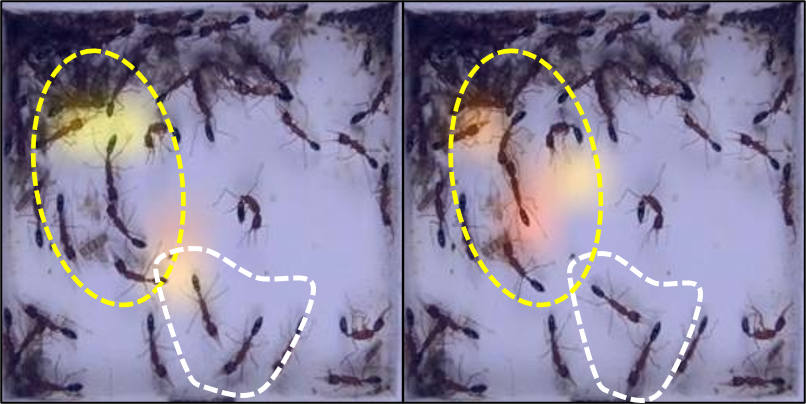}}
        \qquad
        \subfloat[]{\label{fig:duel_non_duel2}
        \includegraphics[width=.4\linewidth]{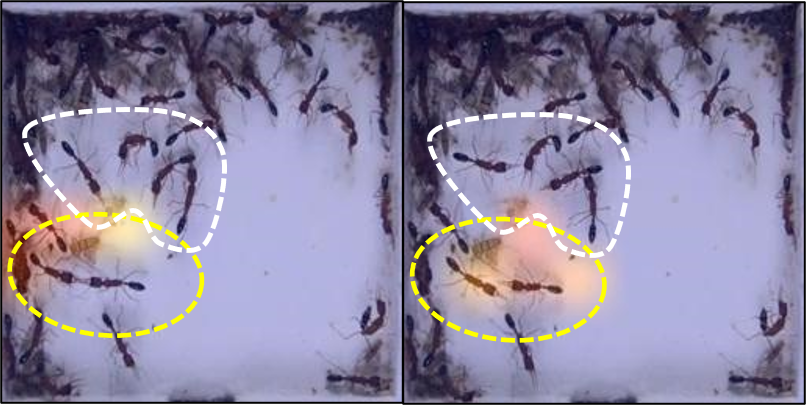}}
        \caption
        {
                Two examples in which \emph{dueling} ants
                are detected (yellow dash line) while other active ones 
                are ignored (white dash line). 
        }
        \label{fig:duel_non_duels}
\end{figure}

\Autoref{fig:centers} displays the heatmaps produced by the initial
application of Grad-CAM with rectifier $\Gamma$. Grad-CAM identifies
that the central area is more critical than the boundaries, and this
general pattern is consistent over time despite changes in ant
behaviors. This visualization indicates that, for the purpose of
identifying changes in colony hierarchical state, the neural network has
learned to ignore interactions near boundaries and instead focuses on
interactions in the center of the area. Although this pattern matches
intuition from human observations of these ants, it is too coarse to
identify important behaviors.

We thus applied a filtered rectifier~$\Gamma'$ to only visualize regions
of the top-$5$\% positive gradients to identify the most dramatic
responses in the generated heatmap to the ant motions, which resulted in
more refined identifications of regions of importance.
\Autorefs{fig:grad_cam_duel_small1}, \ref{fig:grad_cam_duel_small3}, and
\ref{fig:grad_cam_duel_small4} show examples of \emph{dueling}
interactions detected by these highest gradients. Given that the deep
network was not provided coordinates of the ants nor prior behavioral
models of \emph{dueling}, it is not surprising that the highlighted regions do
not precisely identify specific ants in the interactions. Nevertheless,
the network identifies general regions in close proximity to important
behaviors. In particular, in~\autorefs{fig:grad_cam_duel_small3}
and~\ref{fig:grad_cam_duel_small4}, more than two ants were engaged in
\emph{dueling}, but the detection region dynamically moved around them
while they actively participated. These results support that
the trained model has not overfit trivial attributes such as brightness
or contrast of video but learned from ant behaviors themselves.

\Autoref{fig:duel_non_duel1} also shows the case where two duelers are
captured as intended while other active ants who are simply showing
swift turns nearby each other without direct interaction are ignored by
our model. This indicates that the DCNN classifier does not blindly take
any type of movement into account for prediction; only relevant patterns
are prioritized as features to utilize. Similarly,
in~\autoref{fig:duel_non_duel2}, two \emph{dueling} ants are detected
among a group of other non-dueling neighbors that are presenting rapid
changes in motion and orientation. This example also demonstrates the
ability of our trained model to filter out unimportant motion patterns
even when a high degree of motion flow is present.



\section{Summary, Discussion, \& Future Work}
\label{sec:summary}

We have proposed a deep-learning pipeline as a tool to uncover salient
interactions among individuals in a swarm without requiring prior human
knowledge about the behaviors or significant preprocessing effort
devoted to individual tracking and behavioral coding. Our experimental
results show that a trained classifier integrated with Grad-CAM can
localize regions of key individual-scale interactions used by the
classifier to make its colony-scale predictions. Validating our
approach, identified behaviors, such as \emph{dueling}, are the same
behaviors that have been identified previously by human researchers
without the aid of machine learning; however, our classifier discovered
them without any prior guidance from humans. Thus, the library of other
highlighted patterns from our pipeline can be used to generate new
testable hypotheses of individual-to-colony emergence.

Our proposed approach greatly reduces human annotation effort as only
macro-scale, swarm-level annotations are used in training. Significant
effort is currently being used to develop machine-learning models for
the subtask of tracking alone. Our approach suggests that tracking may,
in many cases, be an unnecessary step that wastes both computational and
human resources. Furthermore, our proposed approach reduces the risk of
introducing human bias in the pre-processing of individual-level
observations. Consequently, our example is a model of how human--AI
observational teams can engage in knowledge co-creation~-- each
providing complementary strengths and ultimately realizing the vision of
augmented, as opposed to purely artificial, intelligence.

An important future direction is to further classify the highlighted
patterns automatically discovered by these pipelines. Human behavioral
ecologists can discriminate between peculiar interactions 
(e.g., \emph{dueling}, \emph{dominance biting}, and 
\emph{policing}~\citep{SPSHPL16}) that all may occur 
during the most unstable phases of reproductive hierarchy 
formation in \emph{H.~saltator} ants. Our method may have the ability 
to identify these behaviors, but it does not currently cluster similar identified
patterns together and generate generalizable stereotypes that would be
instructive to human observers hoping to identify these behaviors in
their own future observations. Unsupervised learning methods could be
adopted as a subsequent module to perform clustering and dimensionality
reduction to better communicate common features of clusters, which may
include patterns not yet appreciated by human researchers that are
apparently useful in predicting swarm behavior.

%
%
{\small
\bibliography{bib}
\bibliographystyle{spmpsci}%
}

\end{document}